# Modulation Augmentation and Constellation Shaping for LDPC-Coded QPSK

Heping Wan, Joonyoung Cho, Sandesh Rao Mattu, Jamin Shah, Nishant Mehrotra, Charlie Jianzhong Zhang *Fellow, IEEE*, Robert Calderbank *Life Fellow, IEEE*

***Abstract*—While constellation shaping improves spectral efficiency, its application to quadrature phase-shift keying (QPSK) remains limited. We address this limitation by incorporating a short shaping code into low-density parity-check (LDPC)-coded QPSK systems. Specifically, a subset of LDPC codeword bits is used as shaping bits and mapped by the shaping code to a sparse codeword that enables probabilistic shaping by controlling non-equiprobable selection between lower- and higher-energy QPSK constellations. Because zeros dominate the sparse codeword, lower-energy constellation points are transmitted more frequently, yielding a Gaussian-like symbol distribution. By exploiting the channel-code protection of the shaping bits, we develop receiver algorithms that refine shaping-bit log-likelihood ratios (LLRs) using LDPC decoder feedback. Joint optimization of the shaping codebook, bit-to-symbol mapping and constellations enables the proposed shaping scheme to achieve gains of up to 0.6 and 0.53 dB over 5G New Radio (NR) LDPC-coded QPSK and QAM-16, respectively.**

## I. Introduction

The demand for improved spectral efficiency in next-generation wireless networks has motivated research on constellation shaping as an alternative to uniform signaling [1]–[3]. A fundamental principle of shaping is to transmit lower-energy symbols more frequently than higher-energy ones, reducing average transmit power and approaching the Shannon capacity of the additive white Gaussian noise (AWGN) channel [4].

Probabilistic amplitude shaping (PAS) [1] is a prominent shaping technique. In PAS, a distribution matcher converts uniform information bits into amplitudes drawn from a shaped distribution, which are combined with sign bits and mapped to a higher-order constellation. PAS has demonstrated significant gains for higher-order modulations such as 64-ary and 256-ary quadrature amplitude modulation (QAM). By contrast, QPSK is essential in power-limited scenarios, including cell-edge users and massive Internet-of-Things (IoT) deployments, where higher-order modulations are not viable at low SNRs [5]–[7].

H. Wan, J. Cho, C. Zhang are with Samsung Research America, Plano, Texas, USA (e-mail: heping.wan, joonyoung.cho, jianzhong.z@samsung.com).

J. Shah is with the Department of Electrical and Computer Engineering, University of Minnesota, Twin Cities, MN, USA (e-mail: shah0732@umn.edu)

N. Mehrotra is with the Department of Electrical Communication Engineering at the Indian Institute of Science, Bengaluru (e-mail: nmehrotra@iisc.ac.in)

S. Mattu and R. Calderbank are with the Department of Electrical and Computer Engineering, Duke University, Durham, NC, USA (e-mail: sandesh.mattu, robert.calderbank@duke.edu).

More recently, non-equiprobable signaling via a shaping code over two QPSK constellations has been proposed [8]. In that scheme, shaping bits are encoded into *sparse* codewords that bias transmission toward the lower-energy QPSK constellation. However, the shaping bits are randomly generated and not protected by channel codes, so the receiver cannot refine constellation-selection decisions using side information from the channel-code decoder. This paper addresses this limitation by incorporating a short shaping code into an LDPC-coded QPSK bit-interleaved coded modulation (BICM) system. The main contributions are:

- We propose a shaping scheme in which LDPC codeword bits are partitioned into shaping and non-shaping bits. A shaping code maps the shaping bits to a sparse codeword, which, together with the non-shaping bits, is mapped to an effective 8-ary constellation composed of inner (lower-energy) and outer (higher-energy) QPSK constellations.
- To improve the performance of the proposed shaping scheme, we construct the shaping codebook by omitting the all-zero codeword, which increases the minimum Hamming distance. Furthermore, we introduce a bit-to-symbol mapping rule ensuring that the nearest inner–outer constellation point pair differs only in the shaping-codeword bit, and jointly optimize the powers and phase rotations of the two QPSK constellations to maximize the minimum Euclidean distance.
- We develop two receiver architectures including a serial receiver that performs the demodulation, shaping decoding, and LDPC decoding in a single pass, and an iterative receiver that exchanges extrinsic LLRs among the demodulator, shaping decoder, and LDPC decoder to refine the inner/outer constellation selection.
- The proposed shaping scheme outperforms 5G NR LDPC-coded QPSK and QAM-16 by up to 0.6 dB and 0.53 dB, respectively.

## II. Transmitter Design

### A. Transmitter Overview

As illustrated in Fig. 1, the information bit sequence $\boldsymbol{u}$ is LDPC-encoded, and the resulting codeword $\boldsymbol{c}$ is partitioned into shaping bits $\boldsymbol{s}$ and non-shaping bits $\boldsymbol{v}$. With a $(k_{\mathrm{s}}, n_{\mathrm{s}})$ shaping code, the shaping encoder partitions $\boldsymbol{s}$ into segments of length $k_{\mathrm{s}}$, maps each segment to a length-$n_{\mathrm{s}}$ codeword, and concatenates the codewords to form $\boldsymbol{d}$, which, together with $\boldsymbol{v}$, passes through the shaped modulator. In this paper, the length of $\boldsymbol{s}$ is divisible by $k_s$, and the length of $\boldsymbol{v}$ is twice the length of $\boldsymbol{d}$.

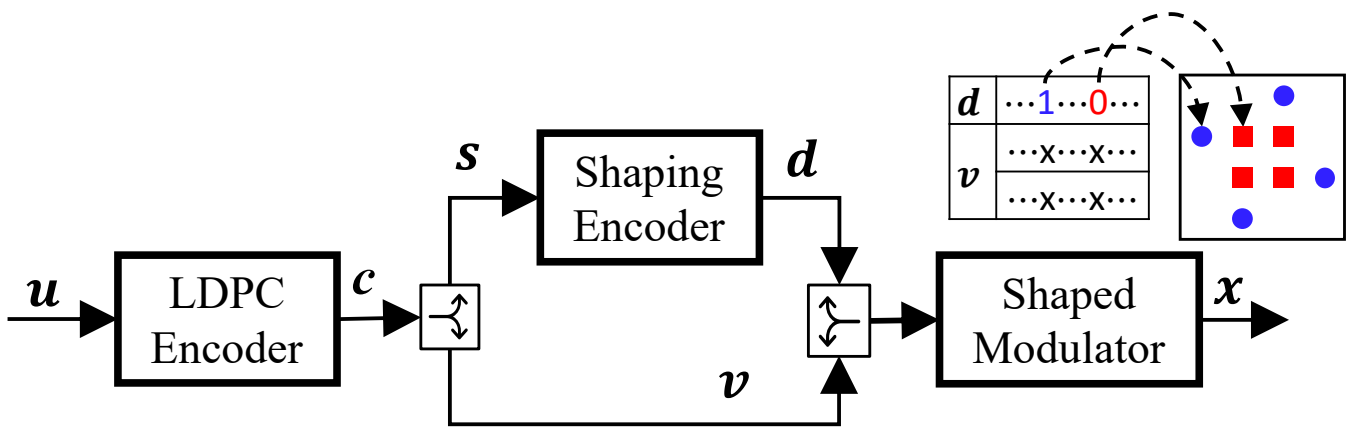

Fig. 1: Block diagram of the proposed transmitter. The inner and outer QPSK constellations, $\mathcal{S}_{\text{inner}}$ and $\mathcal{S}_{\text{outer}}$, are shown as red squares and blue dots, respectively.

The shaped modulator uses two QPSK constellations, $\mathcal{S}_{\text{inner}}$ with lower energy and $\mathcal{S}_{\text{outer}}$ with higher energy. Each bit of $\boldsymbol{d}$ selects the QPSK constellation ("0"$\rightarrow$ inner; "1"$\rightarrow$ outer), and two bits from $\boldsymbol{v}$ select the symbol within that constellation. Thus, each triple $\{b_1, b_2, b_3\}$ with $b_1$ from $\boldsymbol{d}$ and $\{b_2, b_3\}$ from $\boldsymbol{v}$ maps to one point in an effective 8-ary constellation formed by the two QPSK constellations. Because $\boldsymbol{d}$ is sparse and dominated by zeros, inner constellation points are transmitted more frequently, yielding a Gaussian-like symbol distribution.

### B. Shaping Code Design

A classical codebook construction for the $(k_s, n_s)$ shaping code proceeds recursively by Hamming weight [4]. The codebook $\mathcal{D}$ is initialized with the all-zero codeword of length $n_s$. Codewords of increasing weight are then added until $|\mathcal{D}| = 2^{k_s}$.

In this paper, the all-zero codeword is excluded from $\mathcal{D}$. Although the all-zero sequence maximizes inner QPSK constellation usage, its inclusion reduces the minimum Hamming distance. Excluding it increases the minimum Hamming distance and improves shaping-codeword detection reliability.

As an example, a $(4, 16)$ shaping code maps 4 shaping bits to a length-16 codeword. Under the classical construction, the all-zero codeword is included with fifteen weight-one sequences, yielding a minimum Hamming distance of one. In the proposed design, the codebook consists of sixteen weight-one codewords, increasing the minimum Hamming distance to two.

### C. Bit-to-Symbol Mapping Rule

In the proposed design (see Fig. 2a), the closest inner–outer constellation point pair differs *only* in the shaping-codeword bit position. This property aligns the most likely symbol errors with the shaping-bit dimension and improves demodulation reliability. By contrast, under the bit-to-symbol mapping rule in [8], the closest inner–outer pair differs in multiple bit positions, as illustrated in Fig. 2b. Consequently, a single symbol error can induce multiple bit errors.

### D. Constellation Design

The power allocation and phase rotation difference between the inner and outer QPSK constellations are optimized to

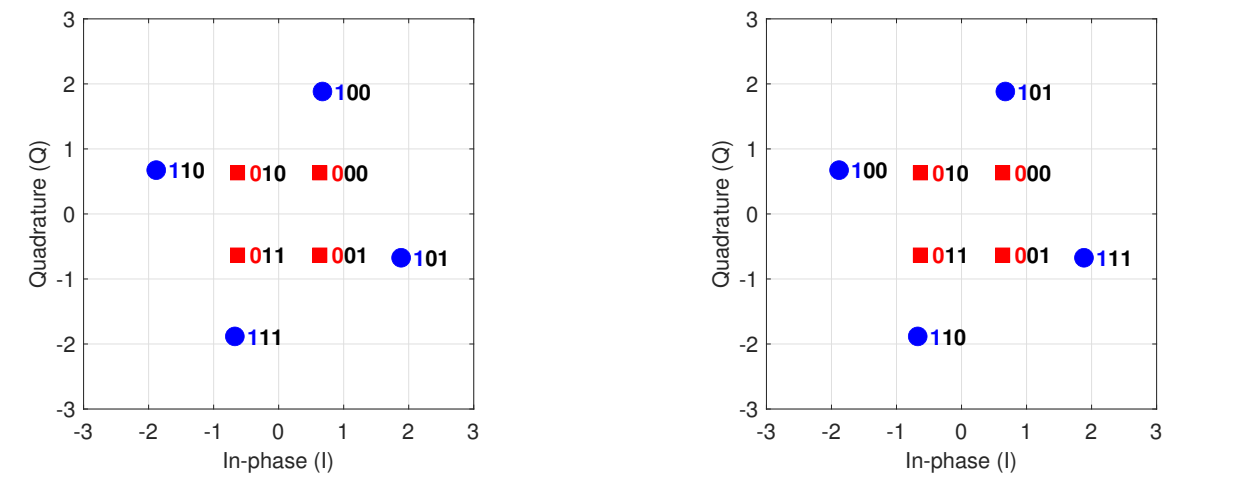

(a) Proposed: One bit difference in the closest inner–outer QPSK point pair

(b) [8]: Two bit difference in the closest inner–outer QPSK point pair

Fig. 2: Bit-to-symbol mapping rules for the effective 8-ary constellation consisting of inner and outer QPSK constellations.

maximize the minimum Euclidean distance subject to the probability distribution induced by the shaping code. Let $P_0$ denote the probability that a shaping codeword bit equals zero, and $E_0$ and $E_1$ denote the per-symbol energy of the inner and outer constellations, respectively. Throughout the paper, the average transmit power of the shaped modulation is normalized to unity,

$$P_0 \times E_0 + (1 - P_0) \times E_1 = 1. \quad (1)$$

A phase rotation $\theta$ is applied to the outer constellation while the phase rotation of the inner constellation is fixed to $\pi/4$.

Fig. 3 illustrates how power allocation affects the minimum Euclidean distance $d_{\text{min}}$. When less power is allocated to the inner QPSK, the distance between two inner-QPSK symbols dominates the minimum distance (see Fig. 3a). Increasing inner-QPSK power shrinks the cross-constellation distance (see Fig. 3b). Fig. 4 shows how the outer-QPSK phase rotation affects the minimum distance.

In this paper, power allocation and phase rotation are jointly optimized via grid search over $\theta \in [0, \pi/2)$ and $E_0 \in (0, 1)$ to maximize the minimum Euclidean distance.[1]

To further improve the performance, the inner and outer constellations are generalized beyond (uniform) QPSK. The resulting constellations in Fig. 5 are obtained using differential evolution [9] with simulated BLER as the optimization metric. Each constellation is constrained to be zero mean to prevent a direct current (DC) component in the baseband waveform and simplify radio-frequency (RF) implementation.

## III. Receiver Design

As shown in Fig. 6, we consider two receiver algorithms for the shaping architecture in Fig. 1. Each comprises a demodulator, a shaping decoder, and an LDPC decoder. The demodulator computes bit LLRs from the received signals and prior probabilities of the shaping-codeword bits. The shaping decoder computes extrinsic LLRs for the demodulator and LDPC decoder by marginalizing over valid codewords in $\mathcal{D}$.

[1]The search space for $\theta$ is limited to $[0, \pi/2)$ due to the symmetry property in QPSK.

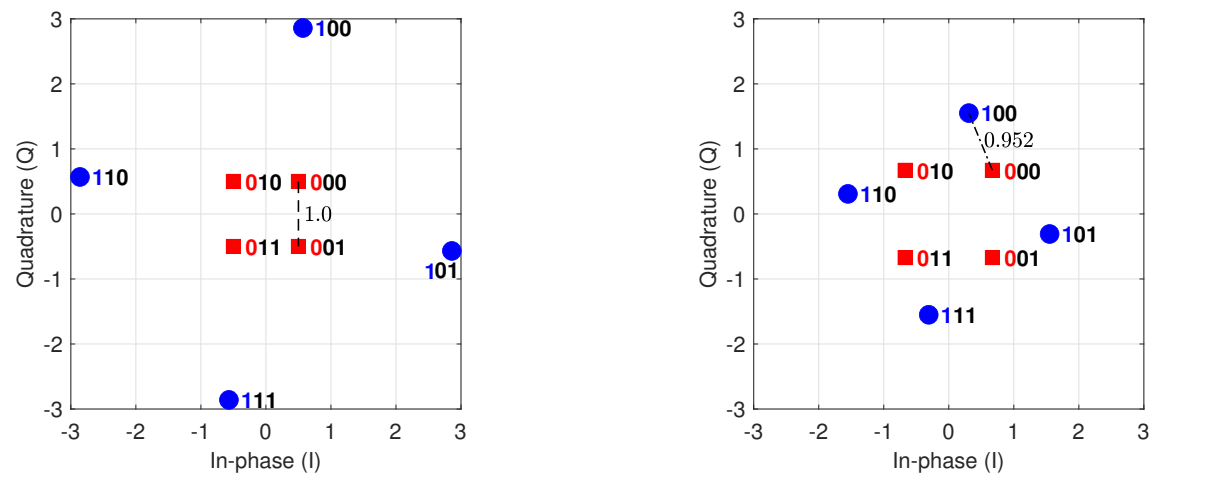

(a) $E_0 = 0.5 \to d_{\min} = 1$ (b) $E_0 = 0.9 \to d_{\min} = 0.952$

Fig. 3: Effect of inner/outer power allocation on the minimum Euclidean distance with $P_0 = 15/16$ and outer-QPSK phase rotation $\theta = 7\pi/16$.

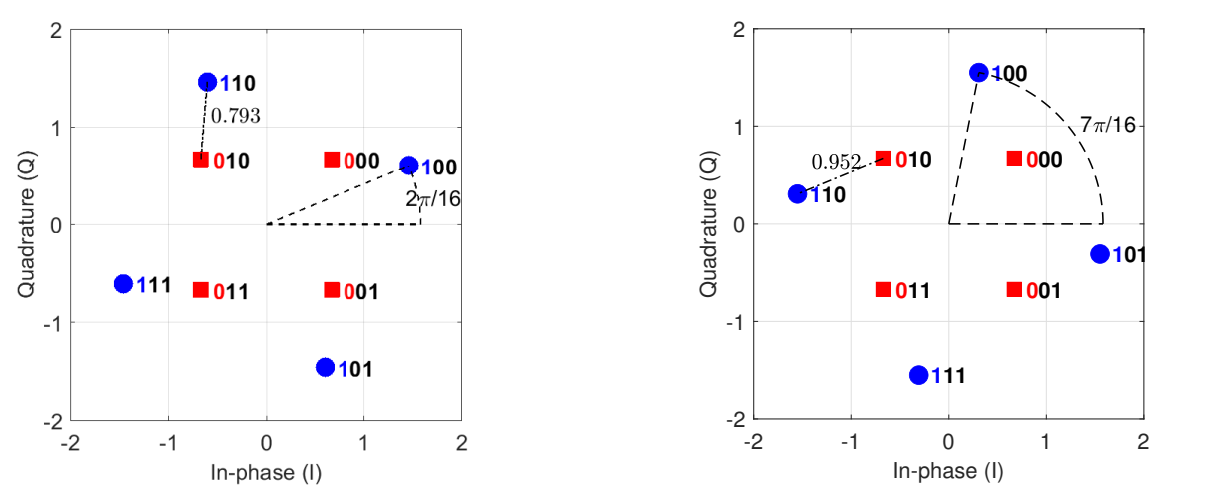

(a) $\theta = 2\pi/16 \to d_{\min} = 0.793$ (b) $\theta = 7\pi/16 \to d_{\min} = 0.952$

Fig. 4: Effect of outer-QPSK phase rotation on the minimum Euclidean distance with $P_0 = 15/16$ and $E_0 = 0.9$.

The LDPC decoder performs belief propagation (BP) using LLRs from the demodulator and shaping decoder.

Let $\boldsymbol{L}_{\boldsymbol{d}}^{(\boldsymbol{y})}$ denote the LLRs for the shaping codeword from the received signals $\boldsymbol{y}$. In both receiver algorithms, these LLRs are computed over the first bit position of the effective 8-ary constellation:

$$L_{b_1} = \log\left(\frac{p(y|b_1=0)}{p(y|b_1=1)}\right) = \log\left(\frac{\sum_{x\in\mathcal{S}_{\text{inner}}} e^{-\frac{|y-x|^2}{N_0}}}{\sum_{x\in\mathcal{S}_{\text{outer}}} e^{-\frac{|y-x|^2}{N_0}}}\right), \quad (2)$$

where $N_0$ denotes the noise power of the AWGN channel.

With a slight abuse of notation, $\boldsymbol{s}$ and $\boldsymbol{d}$ in the following LLR expressions denote a length-$k_s$ shaping-bit sequence and a length-$n_s$ shaping codeword, respectively, rather than the full concatenated sequences in Section II-A. The following computations apply independently to every segment.

### A. Serial Receiver

As shown in Fig. 6a, the demodulation, shaping decoding, and LDPC decoding are performed sequentially in a single pass. The serial receiver proceeds as follows:

1) **Demodulation:** Given prior $P_0$ and the received signals, the LLRs for non-shaping bits $\boldsymbol{L}_{\boldsymbol{v}}$ are computed over the second and third bit positions of the effective 8-ary constellation:

$$L_{b_l} = \log\left(\frac{p(y|b_l=0)}{p(y|b_l=1)}\right), \quad (3)$$

with

$$p(y|b_l=b) = \sum_{x\in\mathcal{S}_{\text{inner}}^{(b_l=b)}} P_0 e^{-\frac{|y-x|^2}{N_0}} + \sum_{x\in\mathcal{S}_{\text{outer}}^{(b_l=b)}} (1-P_0) e^{-\frac{|y-x|^2}{N_0}}, \quad (4)$$

where $\mathcal{S}_{\text{inner}}^{(b_l=b)}$ and $\mathcal{S}_{\text{outer}}^{(b_l=b)}$ denote the subset of the inner and outer QPSK constellations, respectively, when bit $b_l$ is set to $b$ with $l \in \{2,3\}$ and $b \in \{0,1\}$.

2) **Shaping decoding:** Given $\boldsymbol{L}_{\boldsymbol{d}}^{(\boldsymbol{y})}$ from Eq. (2), compute LLRs for shaping bits $\boldsymbol{L}_{\boldsymbol{s}}$. The LLR for the $k$-th shaping bit, where $k \in \{1, 2, \cdots, k_s\}$, is

$$L_{s_k} = \log\left(\frac{\sum_{\tilde{\boldsymbol{d}}\in\mathcal{D}^{(s_k=0)}} p(\boldsymbol{y}|\tilde{\boldsymbol{d}})}{\sum_{\tilde{\boldsymbol{d}}\in\mathcal{D}^{(s_k=1)}} p(\boldsymbol{y}|\tilde{\boldsymbol{d}})}\right),$$

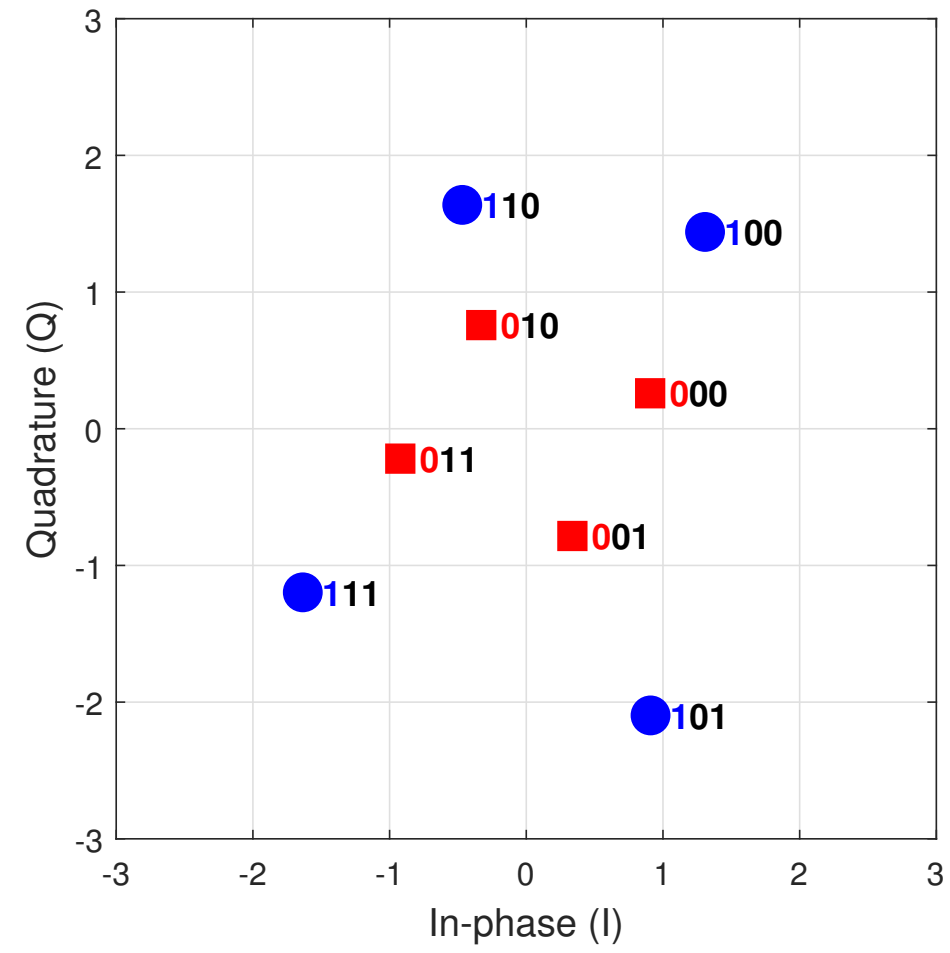

Fig. 5: Optimized inner- and outer- non-uniform 4-ary constellations for the proposed shaping scheme.

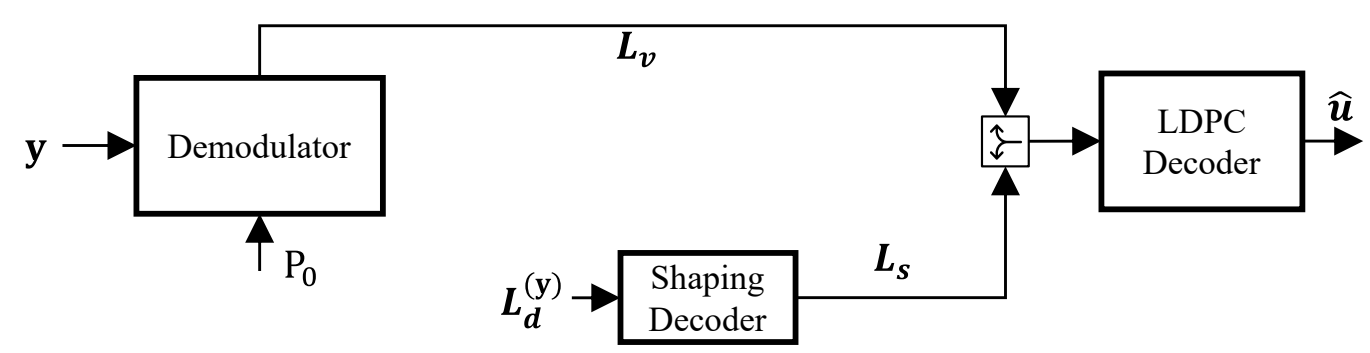

(a) Serial receiver

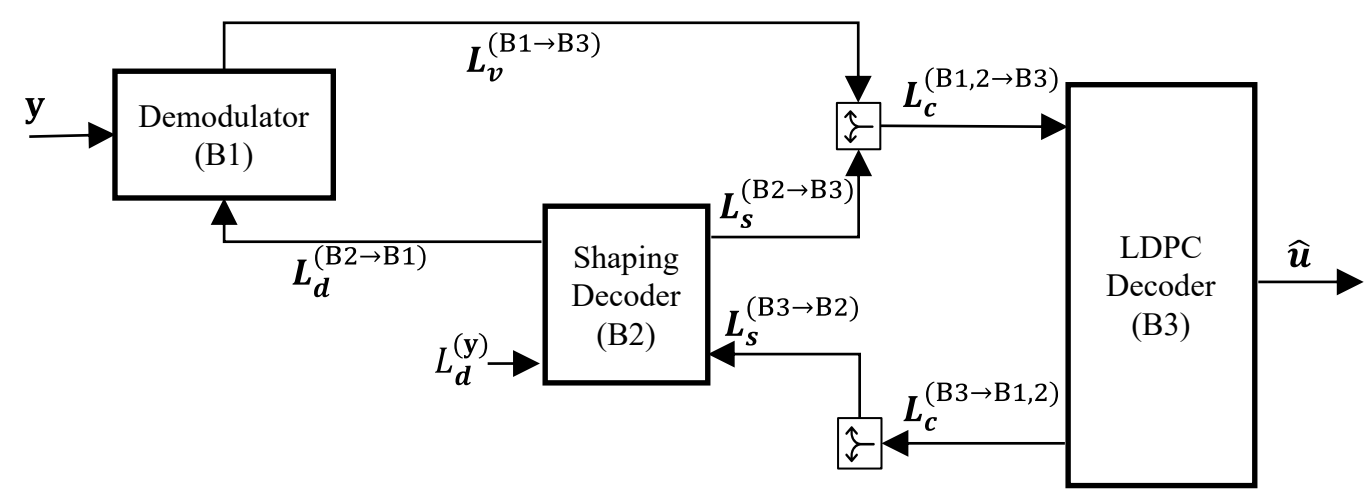

(b) Iterative receiver

Fig. 6: Receiver block diagrams.

where $\mathcal{D}^{(s_k=s)}$ denotes the subset of $\mathcal{D}$ containing shaping codewords associated with the $k-$th shaping bit equal to $s \in \{0,1\}$, and

$$p(\boldsymbol{y}|\tilde{\boldsymbol{d}}) = \prod_{n=1}^{n_s} \frac{e^{(1-\tilde{d}_n) L_{d_n}^{(\boldsymbol{y})}}}{1+e^{L_{d_n}^{(\boldsymbol{y})}}}. \tag{5}$$

3) **LDPC decoding:** Perform BP decoding using the LLRs for shaping bits $\boldsymbol{L_s}$ and non-shaping bits $\boldsymbol{L_v}$ to estimate information bits $\hat{\boldsymbol{u}}$.

### B. Iterative Receiver

As illustrated in Fig. 6b, LLRs are fed back from the shaping decoder to the demodulator and from the LDPC decoder to the shaping decoder. Because the shaping codeword controls selection between the inner and outer QPSK constellations, reliable shaping-bit estimates improve demodulation of the remaining bits, and the refined demodulation in turn benefits LDPC decoding. Per Section IV, the iterative receiver outperforms the serial receiver by approximately 0.25 dB.

For clarity, the demodulator, shaping decoder, and LDPC decoder are denoted as B1, B2, and B3 in Fig. 6b. The algorithm proceeds as follows:

1) **Initialization:** $\boldsymbol{L_d^{(y)}}$ is initialized according to Eq. (2). $\boldsymbol{L_s^{(\text{B3}\to\text{B2})}}$ and $\boldsymbol{L_d^{(\text{B2}\to\text{B1})}}$ are initialized to zeros and $\left(\log \frac{P_0}{1-P_0}\right)$, respectively.

2) **Demodulation:** The calculation follows Eqs. (3) and (4) with $P_0$ replaced by

$$\frac{e^{\boldsymbol{L_d^{(\text{B2}\to\text{B1})}}}}{1+e^{\boldsymbol{L_d^{(\text{B2}\to\text{B1})}}}}.$$

3) **Shaping decoding (forward):** Update LLRs of shaping bits $\boldsymbol{L_s^{(\text{B2}\to\text{B3})}}$ for the LDPC decoder. Given $\boldsymbol{L_d^{(y)}}$ and $\boldsymbol{L_s^{(\text{B3}\to\text{B2})}}$, the LLR for the $k$-th shaping bit is updated as:

$$L_{s_k}^{(\text{B2}\to\text{B3})} = \log\left(\frac{\sum_{\tilde{\boldsymbol{d}}\in\mathcal{D}^{(s_k=0)}} p(\boldsymbol{y}|\tilde{\boldsymbol{d}})p'\left(\mathcal{D}^{-1}(\tilde{\boldsymbol{d}})\right)}{\sum_{\tilde{\boldsymbol{d}}\in\mathcal{D}^{(s_k=1)}} p(\boldsymbol{y}|\tilde{\boldsymbol{d}})p'\left(\mathcal{D}^{-1}(\tilde{\boldsymbol{d}})\right)}\right),$$

where the calculation of $p(\boldsymbol{y}|\tilde{\boldsymbol{d}})$ is the same as Eq. (5). $\mathcal{D}^{-1}(\tilde{\boldsymbol{d}})$ denotes the shaping-bit sequence mapped to $\tilde{\boldsymbol{d}}$ and therefore

$$p'\left(\tilde{\boldsymbol{s}} = \mathcal{D}^{-1}(\tilde{\boldsymbol{d}})\right) = \prod_{k'\in\{1,\cdots,k_s\}\setminus\{k\}} \frac{e^{(1-\tilde{s}_{k'})L_{s_{k'}}^{(\text{B3}\to\text{B2})}}}{1+e^{L_{s_{k'}}^{(\text{B3}\to\text{B2})}}}. \tag{6}$$

4) **LDPC decoding:** Combine LLRs from the demodulator and shaping decoder into $\boldsymbol{L_c^{(\text{B1,2}\to\text{B3})}}$ and perform BP decoding to update LLRs of shaping bits $\boldsymbol{L_s^{(\text{B3}\to\text{B2})}}$ for the shaping decoder. If a stopping condition is met, estimate information bits $\hat{\boldsymbol{u}}$ and terminate.

5) **Shaping decoding (backward):** Update LLRs of shaping-codeword bits $\boldsymbol{L_d^{(\text{B2}\to\text{B1})}}$ for the demodulator. Given $\boldsymbol{L_d^{(y)}}$ and the updated $\boldsymbol{L_s^{(\text{B3}\to\text{B2})}}$, the LLR for the $n$-th shaping-codeword bit, where $n \in \{1,2,\cdots,n_s\}$, is updated as:

$$L_{d_n}^{(\text{B2}\to\text{B1})} = \log\left(\frac{\sum_{\tilde{\boldsymbol{d}}\in\mathcal{D}^{(d_n=0)}} p'(\boldsymbol{y}|\tilde{\boldsymbol{d}})p\left(\mathcal{D}^{-1}(\tilde{\boldsymbol{d}})\right)}{\sum_{\tilde{\boldsymbol{d}}\in\mathcal{D}^{(d_n=1)}} p'(\boldsymbol{y}|\tilde{\boldsymbol{d}})p\left(\mathcal{D}^{-1}(\tilde{\boldsymbol{d}})\right)}\right),$$

where $\mathcal{D}^{(d_n=d)}$ denotes the subset of $\mathcal{D}$ containing shaping codewords with the $n-$th bit equal to $d \in \{0,1\}$,

$$p'\left(\boldsymbol{y}|\tilde{\boldsymbol{d}}\right) = \prod_{n'\in\{1,\cdots,n_s\}\setminus\{n\}} \frac{e^{(1-\tilde{d}_{n'})L_{d_{n'}}^{(\boldsymbol{y})}}}{1+e^{L_{d_{n'}}^{(\boldsymbol{y})}}}, \tag{7}$$

and

$$p\left(\tilde{\boldsymbol{s}} = \mathcal{D}^{-1}(\tilde{\boldsymbol{d}})\right) = \prod_{k=1}^{k_s} \frac{e^{(1-\tilde{s}_k)L_{s_k}^{(\text{B3}\to\text{B2})}}}{1+e^{L_{s_k}^{(\text{B3}\to\text{B2})}}}.$$

The exclusion of $k$ and $n$ from Eqs. (6) and (7) ensures that only extrinsic LLRs are exchanged.

## IV. Simulation Results

### A. Simulation Setup

The proposed scheme employs a $(360, 540)$ 5G NR LDPC code [10], with the first 60 bits of the codeword serving as shaping bits. Combined with a $(4, 16)$ shaping code, this yields a spectral efficiency of 1.5 bits/symbol. Unless stated otherwise, the $(4, 16)$ shaping codebook comprises sixteen weight-one codewords resulting in $P_0 = 15/16$. The baselines include LDPC-coded QPSK and QAM-16 using a $(360, 480)$ and $(360, 960)$ 5G NR LDPC code, respectively, yielding the same spectral efficiency. All simulations assume an AWGN channel and CRC16. BP decoding for LDPC codes employs the layered schedule and syndrome check. For the baseline and the proposed scheme with the serial receiver in Section III-A, up to 10 BP iterations are allowed. For the proposed scheme with the iterative receiver in Section III-B, up to 10 demodulation–decoding iterations are allowed, with one BP iteration per outer iteration.

### B. BLER Performance Comparison

Fig. 7 compares the BLER performance of the proposed scheme with jointly optimized $E_0$ and $\theta$ against 5G NR baselines. With the serial receiver, the proposed scheme achieves 0.35 dB and 0.28 dB gains over LDPC-coded QPSK and 16-QAM, respectively, at BLER $= 10^{-4}$. The iterative receiver further improves these gains to 0.6 dB and 0.53 dB. The shaping gain arises primarily from non-equiprobable inner/outer constellation selection, which produces a Gaussian-like symbol distribution. This gain is further enhanced through optimization of the constellation and shaping codebook, as detailed in the following sections.

### C. Impact of Power Allocation

Fig. 8 evaluates BLER for different power allocated to inner QPSK. The inner- and outer-QPSK phase rotations are fixed to $\pi/4$ and $7\pi/16$, respectively. With $P_0 = 15/16$ induced by the shaping code, the outer-QPSK power follows the average-power constraint in Eq. (1). As illustrated in Fig. 3, assigning too much power to the inner QPSK reduces the minimum distance between inner and outer points, whereas assigning too little power compresses the inner points. An appropriate allocation balances these Euclidean distances. Fig. 8 shows that the best BLER is achieved at $E_0 = 0.8$, which also maximizes the minimum Euclidean distance.

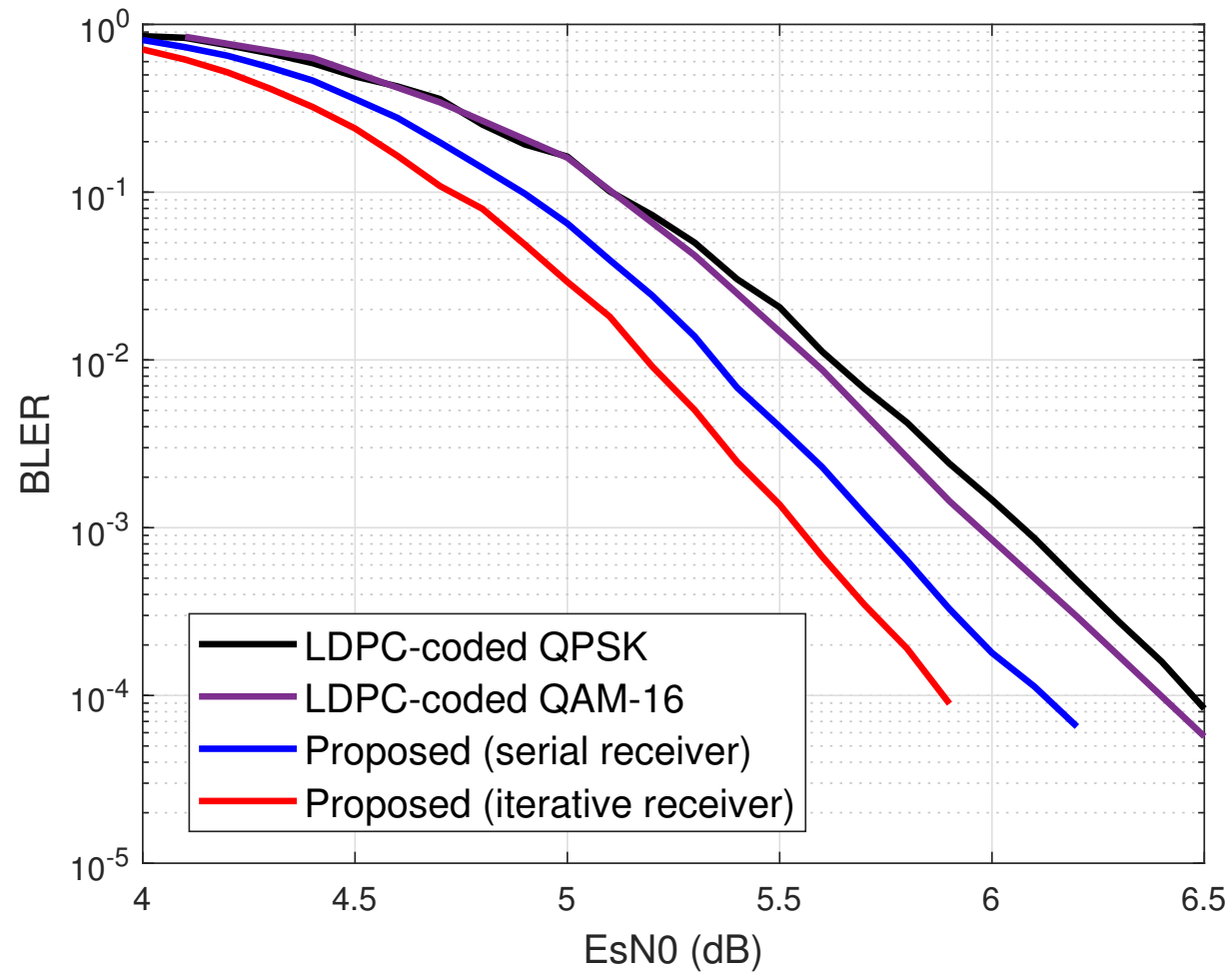

Fig. 7: BLER performance comparison between the proposed shaping scheme with $E_0 = 0.8$ and $\theta = 7\pi/16$ and 5G NR baselines.

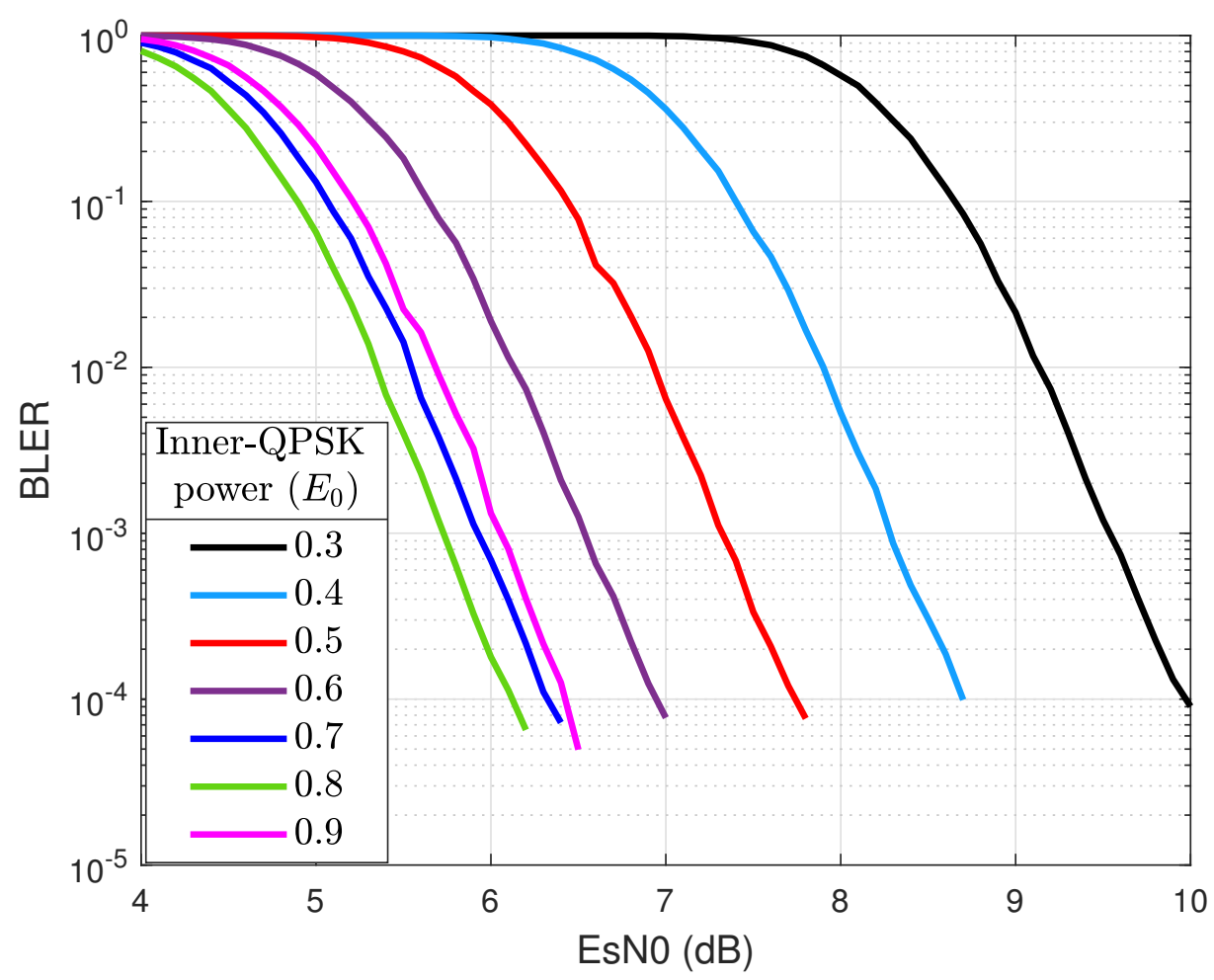

Fig. 8: BLER for various inner-QPSK power allocations.

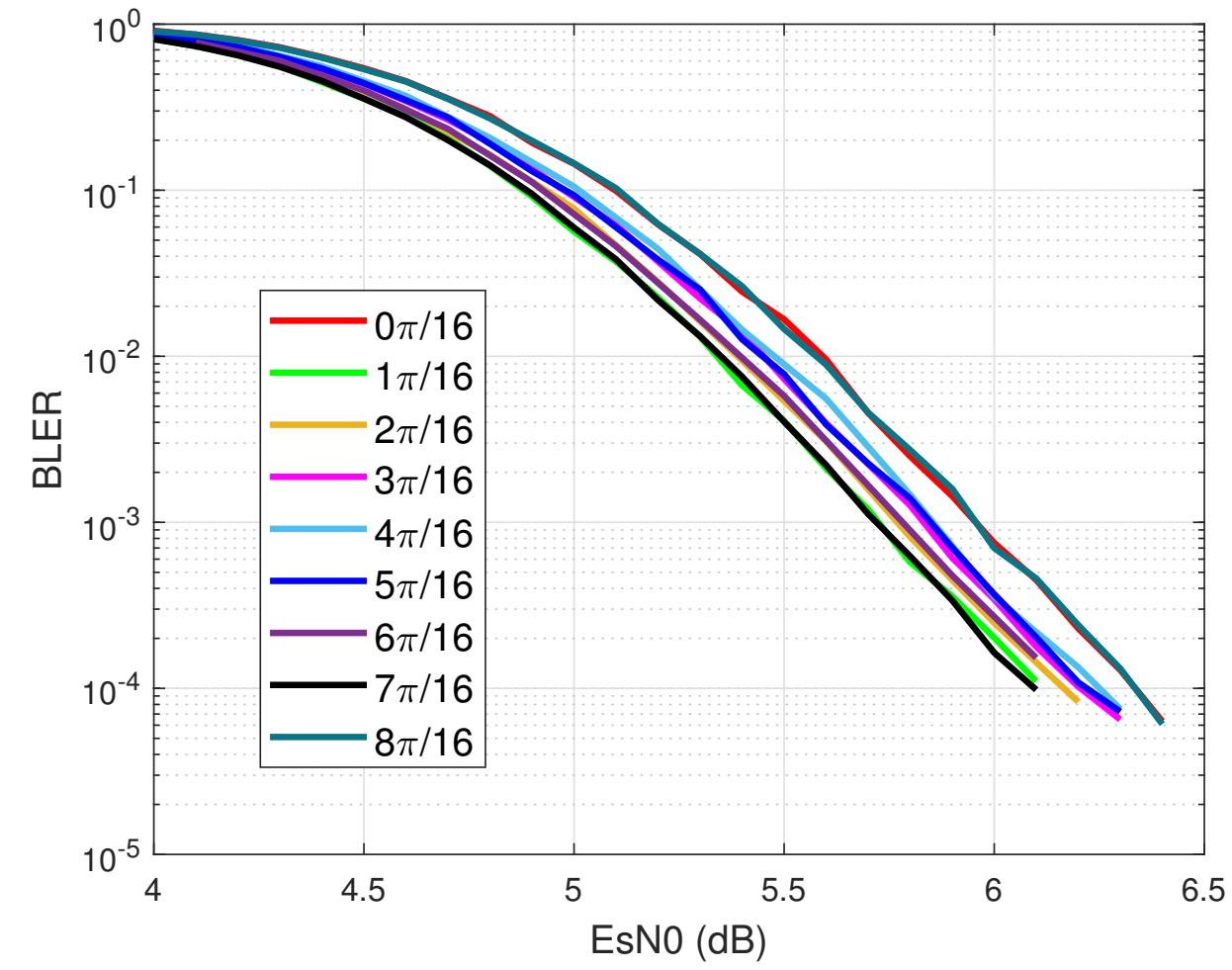

Fig. 9: BLER for various outer-QPSK phase rotations with the inner-QPSK phase rotation fixed to $\pi/4$.

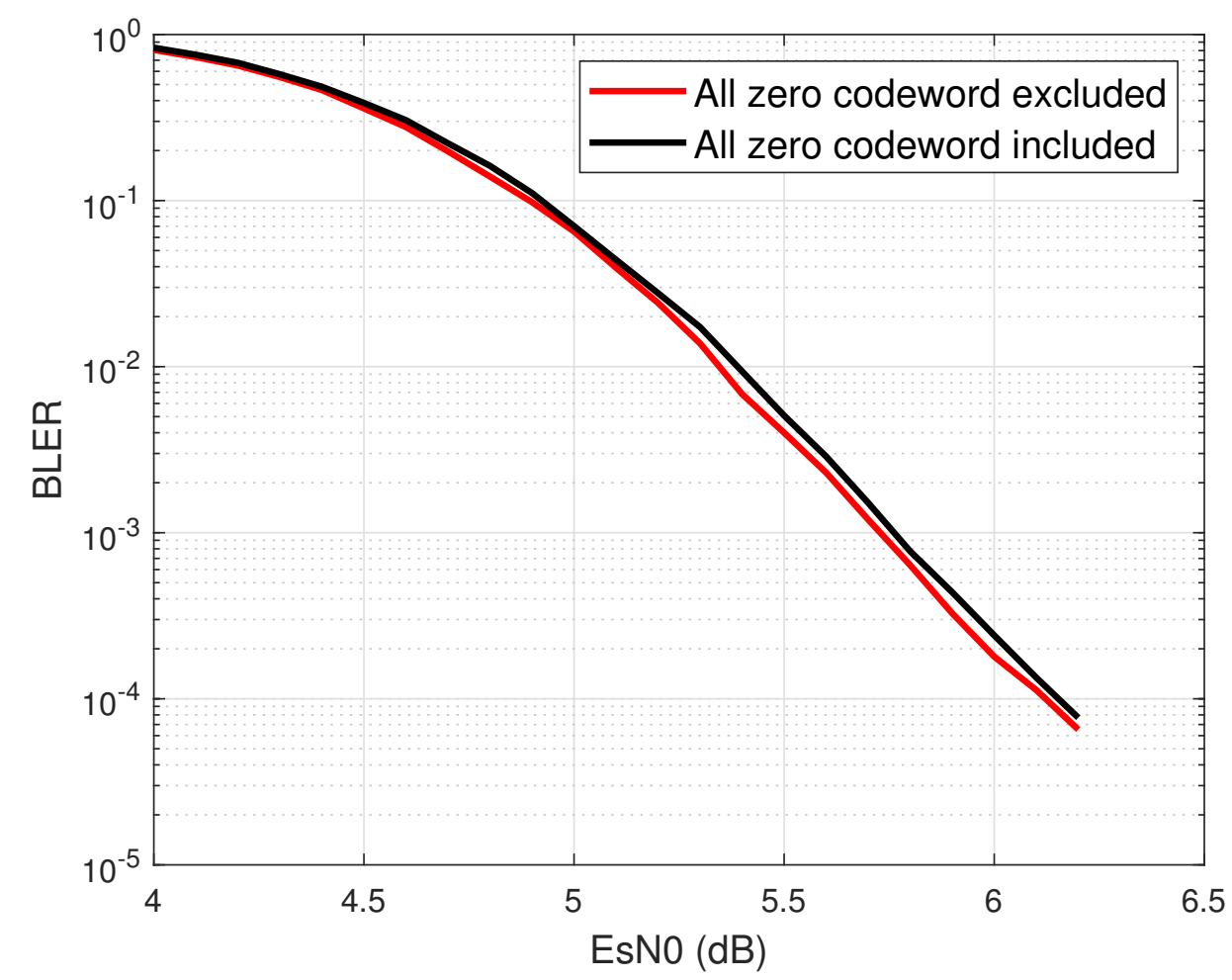

Fig. 10: BLER with and without the all-zero codeword in the $(4, 16)$ shaping codebook.

### D. Impact of Phase Rotation

Fig. 9 compares BLER for different outer-QPSK phase rotations with jointly optimized power allocation. Phase rotations of $1\pi/16$ and $7\pi/16$ (similarly, $0\pi/16$ and $8\pi/16$; $2\pi/16$ and $6\pi/16$; $3\pi/16$ and $5\pi/16$) yield nearly identical BLER, consistent with the symmetry property in QPSK. The best performance occurs at $7\pi/16$ and $\pi/16$, and the corresponding minimum Euclidean distance in Fig. 4 explains their advantage over other choices such as $2\pi/16$.

### E. Impact of Shaping Codebook

Fig. 10 compares the proposed codebook, which excludes the all-zero codeword, with a conventional codebook that includes it. In both cases, $E_0$ and $\theta$ are jointly optimized to 0.8 and $7\pi/16$. Excluding the all-zero codeword increases the minimum Hamming distance from one to two and improves BLER performance.

### F. Impact of Non-Uniform Constellation

Fig. 11 shows a further gain of 0.15 dB from replacing uniform QPSK constellations with non-uniform 4-ary constellations. The optimized constellation in Fig. 5 is used in the non-uniform case. The additional gain reflects the geometric benefit of the optimized non-uniform constellation.

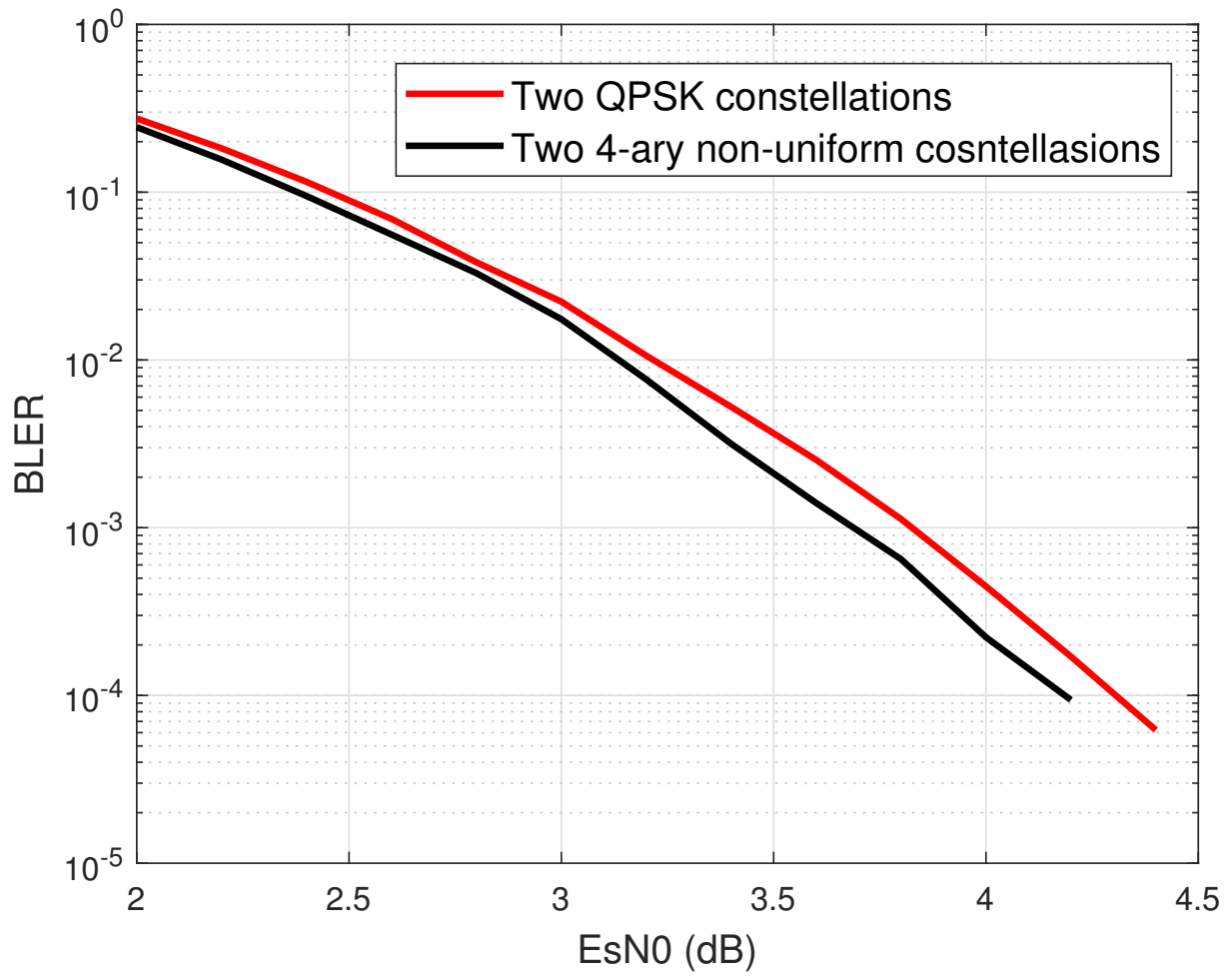

Fig. 11: BLER comparison between uniform QPSK constellations and non-uniform 4-ary constellations.

## V. Conclusion

We proposed a shaping scheme that embeds a short shaping code into an LDPC-coded QPSK BICM system. A portion of LDPC codeword bits serve as shaping bits and are encoded into a sparse codeword that selects between lower- and higher-energy QPSK constellations to induce a Gaussian-like symbol distribution. We optimize the proposed shaping scheme by excluding the all-zero codeword from the codebook to increase the minimum Hamming distance. Furthermore, a dedicated bit-to-symbol mapping rule aligns the dominant symbol errors with the shaping-bit dimension, and the inner/outer power allocation and phase rotation are jointly optimized to maximize the minimum Euclidean distance. The proposed serial and iterative receivers achieve gains of 0.35 dB and 0.6 dB, respectively, over 5G NR LDPC-coded QPSK. Simulation results for various power allocations, phase rotations, and shaping codebooks validate the design.